\documentstyle[prl,aps,floats]{revtex}
\begin{document}
\twocolumn[
\title{Quantum Breathing Mode for Electrons with $1/r^2$ Interaction}
\author{Michael R. Geller and Giovanni Vignale}
\address{Department of Physics, University of Missouri, Columbia, Missouri,
65211}
\date{\today}
\widetext\leftskip=0.10753\textwidth \rightskip\leftskip
\bgroup
\draft
\egroup
\maketitle]

In a recent Letter \cite{Johnson and Quiroga}, Johnson and Quiroga have
obtained several
interesting exact results for electrons with $1/r^2$ interactions in a
two-dimensional
quantum dot. A parabolic confining potential of the form
${\textstyle{1\over 2}} m \omega_0^2 r^2$ is assumed, and the system is
subjected to a uniform perpendicular magnetic field
of strength $B$.
In particular, they have shown that there exists a
collective ``breathing''
mode excitation with frequency
\begin{equation}
\omega = 2 \Omega ,
\label{spectrum}
\end{equation}
where $\Omega^2 \equiv \omega_0^2 + {\textstyle{1 \over 4}} \omega_{\rm c}^2 $,
$\omega_{\rm c} \equiv e B /m c$ is the cyclotron frequency, and
$m$ is the effective mass.
The exact spectrum of the interacting electron system therefore contains an
infinite
ladder of energy levels at integer multiples of $2 \hbar \Omega$.

The purpose of this Comment is to point out that the quantum breathing mode
excitation
spectrum can also be obtained directly from the property of the
many-particle Hamiltonian under a certain scale transformation,
and in a manner that makes
evident the special property of the inverse-square-law interaction for the
quantum breathing mode.
We shall work in the symmetric gauge and write the Hamiltonian as
$ H = T + {\textstyle{1\over 2}} \omega_{\rm c} L_z + V + U, $
where
$T \equiv \sum_n { p_n^2 / 2 m}$, with ${\bf p}_n$  the canonical momentum,
$L_z \equiv \sum_n ({\bf r}_n \times {\bf p}_n) \cdot \hat z$  is the $z$
component of the  canonical angular momentum,
$ V \equiv \sum_n {\textstyle{1 \over 2}} m \Omega^2 r_n^2 $
is the effective field-dependent parabolic confining potential, and
\begin{equation}
U \equiv \sum_{n<n'} {g \over | {\bf r}_n - {\bf r}_{n'} |^\alpha}
\end{equation}
is any {\it power-law} electron-electron interaction.

We first note the properties of $H$ under a
scale transformation
$ O \rightarrow e^{i \lambda S} O e^{-i \lambda S}$
generated by
\begin{equation}
 S \equiv {1 \over 2} \sum_n ( {\bf r}_n \cdot {\bf p}_n
+ {\bf p}_n \cdot {\bf r}_n ).
\end{equation}
This transformation performs a radial displacement of each coordinate by an
amount
proportional to its distance from the origin; that is, it generates a
``breathing''
motion. Under this transformation,
$T \rightarrow T - 2 \lambda T$,
$L_z \rightarrow L_z$,
$V \rightarrow V + 2 \lambda V$,
and
$U \rightarrow U - \alpha \lambda U$,
to first order in $\lambda$. Therefore,
\begin{equation}
H \rightarrow
H + i \lambda [S,H]
= H - 2 \lambda T + 2 \lambda V - \alpha \lambda U .
\label{scaling of H}
\end{equation}
Equation (\ref{scaling of H}), in turn,
may be regarded as an equation of motion for $S$. In fact, noting that
$dV/dt = \Omega^2 S $,
we obtain the operator equation of motion
\begin{equation}
{d^2 V\over d t^2} + (2+\alpha) \Omega^2 V =
\alpha \Omega^2 (H- {\textstyle{1\over 2}} \omega_{\rm c} L_z )
+ (2 - \alpha) \Omega^2 T,
\label{equation of motion}
\end{equation}
which is the same as one would obtain classically.

The breathing mode of the corresponding classical system of point charges
may be obtained from (\ref{equation of motion})
by considering small oscillations about an equilibrium configuration,
where the velocities are zero.
Because $H$ and $L_z$ are constants of the motion, whereas the physical kinetic
energy $T + {1\over 2} \omega_c L_z
+ \omega_c^2 V / 4 \Omega^2$ is zero to  first order in the displacements,
the {\it classical} breathing mode frequency is generally
$\omega = \sqrt{(2 + \alpha) \omega_0^2 + \omega_c^2}.$
For example, the classical breathing frequency of electrons with Coulomb
interactions
($\alpha = 1$)
in a parabolic dot with no magnetic field is
$\sqrt{3} \ \! \omega_0$, a result first obtained by Schweigert and Peeters
\cite{Peeters}.

Quantum zero-point motion, however, generally  modifies the breathing mode and
the other
classical normal modes, by shifting
their frequencies and by giving them a finite lifetime of the order of
$a/R$, with $a$ denoting the Bohr radius and $R$ the radius
of the droplet of charge in the dot.  An exception occurs when $\alpha = 2$: In
this case   $V$ becomes an {\it exact} quantum collective
coordinate with frequency (\ref{spectrum}),
independent of $g$ and $N$.
The collective coordinate $V$ may be separated into a center-of-mass and
relative-coordinate part, $V = V_{\rm cm} + V_{\rm rel}$.
For $\alpha = 2$, it can be shown that each component separately satisifes
a harmonic oscillator equation of motion of the form (\ref{equation of motion})
with frequency $2 \Omega$.
$V_{\rm rel}$ is the collective coordinate corresponding to the breathing
mode discovered by Johnson and Quiroga \cite{Johnson and Quiroga}.

This work was supported by the NSF  Grants No. DMR-9403908 and DMR-9416906. We
gratefully acknowledge the hospitality of the Condensed Matter Theory Group at
Indiana University,  where this work was initiated,   and we thank Allan
MacDonald for  stimulating discussions, and Francois  Peeters for first
drawing our attention to the breathing mode in classical systems.

\end{document}